\documentclass[aps,preprint,prl,color,psfig,epsf]{revtex4}
\usepackage{amsmath}
\usepackage{color}
\usepackage{amsfonts}
\usepackage{times}
\usepackage{epsf}
\usepackage{graphicx}
 \def\dt{\Delta t} 
  
\def\Qab{Q_{\alpha\beta}}

\begin{document}

\title{Lattice Boltzmann simulations of spontaneous flow in active liquid
crystals: the role of boundary conditions}

\author{D. Marenduzzo$^1$, E. Orlandini$^2$, M. E. Cates$^1$, J. M. 
Yeomans$^3$}
\affiliation{$^1$ SUPA, School of Physics, University of Edinburgh,
Mayfield Road, Edinburgh EH9 3JZ, Scotland \\
$^2$ Dipartimento di Fisica and Sezione INFN, Universita' di Padova, Via Marzolo 8, 35131 Padova, Italy \\
$^3$ The Rudolf Peierls Centre for Theoretical Physics, 1 Keble Road, Oxford OX1 3NP, England}

\begin{abstract}
Active liquid crystals or active gels are soft materials which can be
physically realised
e.g. by preparing a solution of cytoskeletal filaments interacting with molecular
motors. We study the hydrodynamics of an active liquid crystal in a slab-like
geometry with various boundary conditions, by solving numerically its equations
of motion via lattice Boltzmann simulations. In all cases we find that active
liquid crystals can sustain spontaneous flow in steady state contrarily to their
passive counterparts, and in agreement with recent theoretical predictions.
We further find that conflicting anchoring conditions at the boundaries
lead to spontaneous flow for any value of the 'activity' parameter, while
with unfrustrated anchoring at all boundaries spontaneous flow only occurs
when the activity exceeds a critical threshold. We finally discuss the
dynamic pathway leading to steady state in a few selected cases.
\end{abstract}

\maketitle


\section{Introduction}

Active viscoelastic gels and active liquid crystals are a
new kind of soft materials
\cite{kruse,joanny,ramaswamy1,ramaswamy2,liverpool}. They differ
from their passive counterparts in the crucial respect that they
continuously burn energy, e.g. from chemical reactions, which
drives them out of thermodynamic equilibrium even in steady state.
Examples of passive viscoelastic fluids are polymeric fluids,
rubber, molten plastics etc. Examples of active gels are
cytoskeletal slabs \cite{bray,cytoskeleton}, biomimetic systems
\cite{beads} or solutions of molecular motors such as dyneins or
myosins and microtubules or actin fibers
\cite{bray,cytoskeleton,peter,beads,nedelec,active_actin_myosin}.
In these solutions the molecular motors interact with actin or
microtubules by exerting forces or torques on them, creating
motile and transient cross-links etc. It is the presence of the
molecular motors, which are able to use chemical energy to perform
mechanical work, which renders these viscoelastic polymeric fluids
`active'. The source of chemical energy in such active
gels typically comes from ATP hydrolysis (there is estimated to be
$\sim 15$ $k_BT$ worth of biochemical energy from each ATP
hydrolysis reaction). It has also been suggested that the
equations of motion of a continuum active gel may explain some
macroscopic aspects of motility in bacterial colonies, such as
swarming (i.e. the collective behaviour of a large number of
bacterial swimmers), elastotaxis etc.
\cite{ramaswamy1,ramaswamy2}.

In general, the modelling of active soft matter presents
novel fundamental challenges to theoreticians, due to the
non-existence of a thermodynamic free energy. Furthermore, the
elastic and rheological properties of these materials may be
anticipated to be strongly affected by their non-equilibrium
nature (`activity'). E.g.,
we now know that spontaneous flow can exist in steady state in
active materials \cite{kruse,joanny}, in sharp contrast to what
happens in their passive counterparts, where any non-zero current
is disallowed in equilibrium by momentum conservation and the
Navier-Stokes equations. Furthermore, many biological gels, such
as actin and neurofilament networks, thicken when sheared
\cite{lubensky}, which is the opposite of the typical behaviour of
viscous polymeric fluids, molten plastics etc., which flow more
easily as the external forcing driving their flow increases.

Our goal here is to study numerically the equations of motion of
an active liquid crystal sample confined between two parallel
plates in a quasi-1D geometry, focussing on the effects of different
anchoring conditions for the director field (the active analog
of the director field), and of the fluid behaviour deep in the
active phase. Numerical methods are needed in this, 
as the insight which is possible to obtain analytically
is limited although important. For instance, it has been
possible to theoretically predict \cite{joanny} by analytic
methods the existence of an activity-induced transition between a
passive phase with no local flow to an `active' phase, in which
the fluid is flowing at steady state. 
However, as the techniques used thus far
essentially allow a stability analysis of the problem in selected
geometries, it is not possible to predict e.g. the flowing and
director field patterns deep in the active phase, away
from the transition to spontaneous flow. Furthermore, using a
numerical method allows us to consider more detailed forms for the
equations of motion. Thus here we extend the vector
polarisation model considered analytically in Ref.
\cite{kruse,joanny}, but a tensorial equivalent (similar to the
one introduced in Ref.\cite{ramaswamy1,ramaswamy2}), which 
allows us to study situations in which the order
parameter is not constant throughout the sample (as occurs in
passive liquid crystals e.g. in shear banding and close to a
defect core).

We will show that the existence of elastic frustration
in the sample rising from the anchoring at the boundaries
is important in determining the steady state solution,
just as it is in selecting director field patterns
of a passive liquid crystal under an imposed shear.
Here we focus on two examples of boundary conditions.
With normal anchoring, we find a transition between passive
and active phase at a (size-dependent) non-zero value of the
`activity' parameter. With conflicting
anchoring (we choose the anchoring leading to the
hybrid aligned nematic -- HAN -- cell in conventional passive
liquid crystals), on the other hand, the behaviour is quite different and
the threshold is zero, i.e. any active liquid crystal in
a HAN geometry spontaneously flows in steady state.
Increasing the activity (going deeper in the active phase),
results in other `transitions'. In the normal anchoring case
we observe banded flow, with the number of bands increasing
as we go deeper into the active phase, while in the HAN cell
geometry a Poiseuille-like flow sets up, and changes
direction upon increasing the activity,
resulting in vastly different elastic deformations of
the sample in steady state.
We also follow the dynamic pathway leading to steady state 
in selected cases and find that transiently the director field
may be transiently asymmetric although it is eventually 
symmetric in steady state. Furthermore we observed some dependence 
of the steady state we find in our simulations on
the initial conditions, hence on the history of the system.

This work is structured as follows. In Section II we introduce the
tensorial model we consider and the equations of motion we want to solve.
We also briefly show how the lattice Boltzmann algorithm of Ref. \cite{lblc}
may be generalised to solve these equations of motion. In Section III
we report our results for the flow and director field patterns in
steady state in the case of normal (Section III A) and conflicting
(Section III B) boundary conditions at the sample walls. 
In both cases we assume translational invariance parallel to the
walls so that the numerical problem is effectively 
one dimensional. We also
discuss the dynamics leading to steady state in a selected case with
normal anchoring (Section III C). 
Finally, Section IV contains our conclusions.


\section{Model and Methods}

We introduce a Landau-de Gennes free energy to describe the
equilibrium physics of the active liquid crystal.
This is a function of a tensorial order parameter,
$Q_{\alpha\beta}$, whose largest eigenvalue gives the strength of
local order. The free energy ${\cal F}$ is a sum of two terms,
${\cal F}_1+{\cal F}_2$.
The first one is a bulk contribution,
\begin{eqnarray}
{\cal F}_1=\frac{A_0}{2}(1 - \frac {\gamma} {3}) Q_{\alpha \beta}^2 -
          \frac {A_0 \gamma}{3} Q_{\alpha \beta}Q_{\beta
          \gamma}Q_{\gamma \alpha}
+ \frac {A_0 \gamma}{4} (Q_{\alpha \beta}^2)^2,
\label{eqBulkFree}
\end{eqnarray}
while the second, ${\cal F}_2=K\left(\partial_\gamma 
Q_{\alpha \beta}\right)^2/2$
is a distortion term \cite{degennes}.
Hereafter Greek indices denote Cartesian components and
summation over repeated indices is implied.
$K$ is an elastic constant. We have taken the one elastic
costant approximation and neglected the spontaneous splay
coefficient, which is equivalent to having infinitely strong
anchoring at the boundaries. Adding a spontaneous splay does
not change the physics qualitatively as argued in \cite{joanny}.
In the free energy equation, $A_0$ is a constant and $\gamma$
controls the magnitude of order.
The anchoring of the director field on the boundary surfaces
is along $\hat{x}$ or  $\hat{x}$, see Fig. 1.

The equation of motion for {\bf Q} is then \cite{beris}
\begin{equation}
(\partial_t+{\vec u}\cdot{\bf \nabla}){\bf Q}-{\bf S}({\bf W},{\bf
  Q})= \Gamma {\bf H}+\lambda \Delta\mu {\bf Q}
\label{Qevolution}
\end{equation}
where $\Gamma$ is a collective rotational diffusion constant,
$\Delta\mu$ is the chemical potential gained via
ATP hydrolysis \cite{kruse},
and $\lambda$ is an `activity' parameter
of the liquid crystal.

Eq. (\ref{Qevolution}) is valid to first order in 
$\Delta \mu$ \cite{kruse}.
The first term on the left-hand side of Eq. (\ref{Qevolution})
is the material derivative describing a
quantity advected by a fluid with velocity ${\vec u}$. This is
generalized for rod-like molecules by a second term
\begin{eqnarray}\label{S_definition}
{\bf S}({\bf W},{\bf Q})
& = &(\xi{\bf D}
+{\bf \omega})({\bf Q}+{\bf I}/3)\\ \nonumber
&+& ({\bf Q}+
{\bf I}/3) (\xi{\bf D}-{\bf \omega})
- 2\xi({\bf Q}+{\bf I}/3){\mbox{Tr}}({\bf Q}{\bf W})
\end{eqnarray}
where Tr denotes the tensorial trace, while
${\bf D}=({\bf W}+{\bf W}^T)/2$ and
${\bf \omega}=({\bf W}-{\bf W}^T)/2$
are the symmetric part and the anti-symmetric part respectively of the
velocity gradient tensor $W_{\alpha\beta}=\partial_\beta u_\alpha$.
The constant $\xi$ depends on the molecular
details of a given liquid crystal. In our context increasing $\xi$
tends to disfavour tumbling and favour aligning of the director in
the material.
The molecular field ${\bf H}$ in Eq. \ref{Qevolution}
is given by ${\bf H}= -{\delta {\cal F} \over \delta {\bf Q}}+({\bf
    I}/3) Tr{\delta {\cal F} \over \delta {\bf Q}}$.

The fluid velocity, $\vec u$, obeys the continuity equation and
the Navier-Stokes equation,
\begin{eqnarray}\label{navierstokes}
\rho(\partial_t+ u_\beta \partial_\beta)
u_\alpha & = & \partial_\beta (\Pi_{\alpha\beta})+
\eta \partial_\beta(\partial_\alpha
u_\beta + \partial_\beta u_\alpha\\ \nonumber
& + & (1-3\partial_\rho
P_{0}) \partial_\gamma u_\gamma\delta_{\alpha\beta}),
\end{eqnarray}
where $\rho$ is the fluid density and $\eta$ is an isotropic
viscosity, and $\Pi_{\alpha\beta}=\Pi^{\rm passive}_{\alpha\beta}+
\Pi^{\rm active}_{\alpha\beta}$.
$\Pi^{\rm passive}_{\alpha\beta}$ is the stress tensor necessary to describe
ordinary liquid crystal hydrodynamics:
\begin{eqnarray}
\Pi^{\rm passive}_{\alpha\beta}= &-&P_0 \delta_{\alpha \beta} +2\xi
(Q_{\alpha\beta}+{1\over 3}\delta_{\alpha\beta})Q_{\gamma\epsilon}
H_{\gamma\epsilon}\\\nonumber
&-&\xi H_{\alpha\gamma}(Q_{\gamma\beta}+{1\over
  3}\delta_{\gamma\beta})-\xi (Q_{\alpha\gamma}+{1\over
  3}\delta_{\alpha\gamma})H_{\gamma\beta}\\ \nonumber
&-&\partial_\alpha Q_{\gamma\nu} {\delta
{\cal F}\over \delta\partial_\beta Q_{\gamma\nu}}
+Q_{\alpha \gamma} H_{\gamma \beta} -H_{\alpha
 \gamma}Q_{\gamma \beta} .
\label{BEstress}
\end{eqnarray}
$P_0$ is a constant in the simulations reported here.
The active term is given by
$\Pi^{\rm active}_{\alpha\beta}=\zeta \Delta\mu Q_{\alpha\beta}$,
where $\zeta$ is another `activity' parameter. This equation is
also valid at first order in $\Delta \mu$ \cite{kruse}.

The model above reduces to the Beris-Edwards model for liquid crystal
hydrodynamics for $\lambda=\zeta=0$. For uniaxial active liquid crystals, the
active liquid crystal director field $\vec n$ is well defined through
$Q_{\alpha\beta}=3q\left(n_{\alpha}
n_{\beta}-\delta_{\alpha\beta}/3\right)/2$, where $q$ is the largest
eigenvalue of $Q_{\alpha\beta}$. In this limit our model can be
shown to reduce to the vectorial model considered in
\cite{kruse,joanny}.

We solve Eqs. \ref{Qevolution} and \ref{navierstokes} by using a
Lattice Boltzmann algorithm. Our algorithm is a generalisation of
the one introduced in Refs. \cite{lblc} and successfully used to
study a number of problems in the hydrodynamics and rheology of a
passive liquid crystal. We now review it briefly, following
the treatment in Ref. \cite{lblc}.

Lattice Boltzmann algorithms to solve the Navier-Stokes
equations of a simple fluid, are defined in terms of a single set of partial
distribution functions, the scalars $f_i (\vec{x})$, that sum on each lattice
site $\vec{x}$ to give the density.
For liquid crystal hydrodynamics, this
must be supplemented by a second set, the symmetric traceless tensors
${\bf G}_i (\vec{x})$, that are related to the tensor order parameter
${\bf Q}$. Each $f_i$, ${\bf G}_i$ is associated with a lattice
vector ${\vec e}_i$ \cite{lblc}.  We choose a 15-velocity model on the cubic
lattice with lattice vectors:
\begin{eqnarray}
\vec {e}_{i}^{(0)}&=& (0,0,0)\\
\vec {e}_{i}^{(1)}&=&(\pm 1,0,0),(0,\pm 1,0), (0,0,\pm 1)\\
\vec {e}_{i}^{(2)}&=&(\pm 1, \pm 1, \pm 1).
\label{latvects}
\end{eqnarray}
The indices, $i$, are ordered so that $i=0$ corresponds to
$\vec {e}_{i}^{(0)}$, $i=1,\cdots 6$ correspond to the $\vec {e}_{i}^{(1)}$
set and $i=7,14$ to the $\vec {e}_{i}^{(2)}$ set.

Physical variables are defined as moments of the distribution function:
\begin{equation}
\rho=\sum_i f_i, \qquad \rho u_\alpha = \sum_i f_i  e_{i\alpha},
\qquad {\bf Q} = \sum_i {\bf G}_i.
\label{eq1}
\end{equation}

The distribution functions evolve in a time step $\Delta t$ according
to
\begin{equation}
f_i({\vec x}+{\vec e}_i \Delta t,t+\Delta t)-f_i({\vec x},t)=
\frac{\Delta t}{2} \left[{\cal C}_{fi}({\vec x},t,\left\{f_i
\right\})+ {\cal C}_{fi}({\vec x}+{\vec e}_i \Delta
t,t+\Delta
t,\left\{f_i^*\right\})\right]
\label{eq2}
\end{equation}
\begin{equation}
{\bf G}_i({\vec x}+{\vec {e}}_i \Delta t,t+\Delta t)-{\bf G}_i({\vec
x},t)=\frac{\Delta t}{2}\left[ {\cal C}_{{\bf G}i}({\vec
x},t,\left\{{\bf G}_i \right\})+
                {\cal C}_{{\bf G}i}({\vec x}+{\vec {e}}_i \Delta
                t,t+\Delta t,\left\{{\bf G}_i^*\right\})\right]
\label{eq3}
\end{equation}
This represents free streaming with velocity ${\vec e}_i$ followed by a
collision step which allows the distribution to relax towards
equilibrium.
The ${\cal C}$'s are collision operator (detailed below) while
$f_i^*$ and ${\bf G}_i^*$ are first order approximations to
$f_i({\vec {x}}+{\vec {e}}_i \dt,t+\dt)$ and ${\bf G}_i({\vec {x}}+{\vec {e}}_i \dt,t+\dt)$
respectively.
They are obtained by using $\Delta t\, {\cal C}_{{\bf f}i}({\vec
x},t,\left\{{\bf f}_i \right\})$ on the right hand side of Eq. (\ref{eq2}) and
a similar substitution in Eq. (\ref{eq3}).
Discretizing in this way, which is similar to a predictor-corrector
scheme, has the advantages that lattice viscosity terms are eliminated
to second order and that the stability of the scheme is improved.

The collision operators are taken to have the form of a single
relaxation time Boltzmann equation, together with a forcing term
\begin{equation}
{\cal C}_{fi}({\vec {x}},t,\left\{f_i \right\})=
-\frac{1}{\tau_f}(f_i({\vec {x}},t)-f_i^{eq}({\vec {x}},t,\left\{f_i
\right\}))
+p_i({\vec {x}},t,\left\{f_i \right\}),
\label{eq4}
\end{equation}
\begin{equation}
{\cal C}_{{\bf G}i}({\vec x},t,\left\{{\bf G}_i
\right\})=-\frac{1}{\tau_{\bf G}}({\bf G}_i({\vec x},t)-{\bf
G}_i^{eq}({\vec x},t,\left\{{\bf G}_i \right\}))
+{\bf M}_i({\vec x},t,\left\{{\bf G}_i \right\}).
\label{eq5}
\end{equation}

The form of the equations of motion
follow from the choice of the moments of the equilibrium distributions
$f^{eq}_i$ and ${\bf G}^{eq}_i$ and the driving terms $p_i$ and
${\bf M}_i$. The distributions $f_i^{eq}$ are constrained by
\begin{equation}
\sum_i f_i^{eq} = \rho,\qquad \sum_i f_i^{eq} e_{i \alpha} = \rho
u_{\alpha}, \qquad
\sum_i f_i^{eq} e_{i\alpha}e_{i\beta} = -\Pi_{\alpha\beta}^{\rm passive}+\rho
u_\alpha u_\beta
\label{eq6}
\end{equation}
where the zeroth and first moments are chosen to impose conservation of
mass and momentum. The second moment of $f^{eq}$ controls the symmetric
part of the stress tensor, whereas the moments of $p_i$
\begin{equation}
\sum_i p_i = 0, \quad \sum_i p_i e_{i\alpha} = \partial_\beta
\tau_{\alpha\beta},\quad \sum_i p_i
e_{i\alpha}e_{i\beta} = 0
\label{eq7}
\end{equation}
impose the antisymmetric part of the stress tensor,
which we have called $\tau_{\alpha\beta}$.
For the equilibrium of the order parameter distribution we choose
\begin{equation}
\sum_i {\bf G}_i^{eq} = {\bf Q},\qquad \sum_i
{\bf G}_i^{eq} {e_{i\alpha}} = {\bf Q}{u_{\alpha}},
\qquad \sum_i {\bf G}_i^{eq}
e_{i\alpha}e_{i\beta} = {\bf Q} u_\alpha u_\beta .
\label{eq8}
\end{equation}
This ensures that the order parameter
is convected with the flow. Finally the evolution of the
order parameter is most conveniently modeled by choosing
\begin{equation}
\sum_i {\bf M}_i = \Gamma {\bf H}({\bf Q})
+{\bf S}({\bf W},{\bf Q}) \equiv {\bf \hat{H}}, \qquad
\qquad \sum_i {\bf M}_i {e_{i\alpha}} = (\sum_i {\bf M}_i)
{u_{\alpha}}
\label{eq9}
\end{equation}
which ensures that the fluid minimises its free energy at equilibrium.

Conditions (\ref{eq6})--(\ref{eq9}) are satisfied,
by writing the equilibrium distribution functions and
forcing terms as polynomial expansions in the velocity.
The coefficients in the expansion are (in general non-uniquely)
determined by the requirements that these constraints are
fulfilled (see Ref. \cite{lblc} for details). 

The active contributions
then alter the constraints on the second moment of the
$f_i$'s and the first moment of the ${\bf G}_i$'s distribution
functions as follows:
\begin{equation}
\sum_i {\bf M}_i = {\bf \hat{H}}+\lambda\Delta\mu {\bf Q}, \qquad
\sum_i f_i^{eq} e_{i\alpha}e_{i\beta} = -\Pi_{\alpha\beta}^{\rm passive}+\rho
u_\alpha u_\beta-\zeta\Delta\mu\Qab.
\label{activity_changes}
\end{equation}
Alternatively, we can input the derivative of the pressure tensor
as a body force (as in Eq. \ref{eq7}). This scheme rigorously ensures
no spurious velocity in equilibrium \cite{lblc,wagner} and has been employed 
in the following.

\section{Results}

In this Section we report the results obtained with lattice
Boltzmann simulations of the dynamic behaviour of a slab of active
liquid crystal, sandwitched between two fixed plates at a mutual distance $L$
along the $y$ axis (see Fig. \ref{setup} for axis definition). As
a result our system is effectively one dimensional (the velocity
and the active liquid crystal director fields depend on the velocity gradient
coordinate, $y$, only) We shall consider different kinds of
boundary conditions, and demonstrate that these affect the
hydrodynamics of the fluid considerably.

The model parameters in the simulations reported here were
typically fixed as follows: $A_0=0.1$, $L=100$, $K=0.04$, $\gamma=3$,
$\xi=0.7$, $\Gamma\sim 0.34$, while the relaxation time of
the lattice Boltzmann evolution of the density distribution
function (see Ref. \cite{lblc} for details) is $0.7$.
All preceding values are in simulation units. These can be
straightforwardly related to a set of physical units by
considering that one space and time simulation units correspond
to $0.02$ $\mu$m and $0.067$ $\mu$s respectively (for how to
do this mapping see Ref. \cite{lblc}).
One can thus map our sample onto a physical slab of active liquid
crystal of thickness $10$ $\mu$m, while the parameters of
the liquid crystalline fluids are $K=16.2$ pN,
$\gamma=1$ Poise ($\gamma$ is the rotational viscosity \cite{degennes}),
while the ratio between the Leslie viscosities $\alpha_3/\alpha_2\sim 0.08$.
Note that a positive value of $\alpha_3/\alpha_2$ means that the liquid crystal in
the passive phase is flow aligning \cite{degennes}; thus
the parameter choice made here corresponds to
a flow aligning liquid crystal in the passive ($\zeta=0$) phase.
We have further fixed
$\lambda=0$ in our simulations -- as this parameter has been shown to not
affect the existence of a transition between a passive quiescent
and an active phase with spontaneous flow \cite{joanny}.
The values of the remaining `activity' parameter $\zeta$
were on the other hand systematically varied.

In what follows the director field (also called polarisation field in
the context of active gels in Ref. \cite{joanny}) remains in the
$xy$ plane in the vast majority of cases. It is therefore useful to
introduce an orientation angle $\theta$, as the angle between the director
field and the positive direction along the $y$ axis.

Note that, unless specified otherwise, we always initialised the system with
zero velocity field and with a director field corresponding to
the stable equilibrium one in the passive phase. However,
as this initial state is generally metastable, and as our lattice
Boltzmann simulations do not include noise, we introduced a small
perturbation -- typically in the form of a small deviation in the orientation
angle at $y=L/2$. This is enough to allow the system to escape 
from the metastable passive phase in which it is initialised.

\subsection{Active liquid crystals with normal anchoring}

We first consider an active liquid crystal with normal anchoring at its
boundaries (Fig. 1a).

For values of $\zeta$ smaller than $\zeta_c\sim 0.0015$ the liquid
crystal remains in the passive phase, whereas for $\zeta>\zeta_c$
there is a finite flow along the $x$ direction (Fig. \ref{setup}),
i.e. the liquid crystal is in an active phase. Fig. \ref{N1} shows the
director and velocity profiles first encountered in the active
phase upon increasing the value of $\zeta$. There is a net primary
flow along the $x$ direction, which exhibits a sharp maximum at
the channel center, and is quite different from e.g. the common
parabolic Poiseuille velocity profile which may be externally
induced in a passive liquid crystal by exerting a pressure
difference or a body force on it. The spontaneous flow
symmetrically deforms the director profile and two 'bands'
develop, in which the orientation angle $\theta$ attains opposite
values (close to that of the Leslie angle \cite{degennes} which is
imposed by the local shear rate introduced by the spontaneous
flow). These band flatten when the value of $\zeta$ is increased
-- up to a second critical value, which for the parameters we use
here is $\sim 0.0015$.

A further increase in $\zeta$ leads to a qualitatively different
behaviour, as shown in Fig. \ref{N2}. The velocity profile changes and
now has a sinusoidal shape, so that it is zero at the centre of the channel
and there is no net flow. For values of $\zeta$ close to this second crossover
the director field exhibits an out-of-plane instability, much as conventional
(yet flow tumbling) liquid crystals under strong shear. For large $\zeta$
the director field is well aligned along the shear direction (Fig. \ref{N2}).

\subsection{Active liquid crystal with conflicting anchoring}

We now come to the case of an active liquid crystal with
the director field along $y$ at
$y=L$ and along $x$ at $y=0$. In the passive case the
liquid crystal responds to this conflicting anchoring by
adopting a uniform (in the one elastic constant
approximation employed here) splay-bend deformation
fully contained in the $xy$ plane.

Fig. \ref{HAN1} shows the director and velocity fields
at steady state for small values of the activity $\zeta$. Contrarily
to the normal anchoring case, we find no critical non-zero
value of $\zeta$ for which the system is passive. Formally
this can be understood as in the passive steady state there
are gradients of ${\bf Q}$ which set up a non-trivial shear
rate so that the passive solution is always unstable (for
$\zeta>0$). For small values of $\zeta$, the active gel
expands the region which is aligned along $x$, while the flow field
is asymmetric at steady state.

Fig. \ref{HAN2} proves that also in this HAN geometry further
`transitions' are encountered when going deeper into the active
phase by increasing $\zeta$. For $\zeta>\zeta^*\sim 0.0003$, the
sign of the flow in steady state reverses. Increasing the activity
also leads to a more symmetric, Poiseuille-like flow. The change
in orientation of the fluid flow relative to the director
field leads to a strikingly different elastic deformation which
ends up with a point with $\theta=0$ (director completely
along the velocity gradient direction) within the sample. The
director profile can be understood as due to the fluid flow in
steady state, as it resembles the elastic distortions in steady
state found with clockwise and anti-clockwise HAN cells under a
constant shear flow in Ref. \cite{distorted}. It would be of
interest to develop a semi-analytic theory which accounts for the
presence of extra `transitions', both in the HAN and in the normal
anchoring geometry, when the activity is increased.

\subsection{Dynamic pathway to steady state}

We now consider the dynamic pathway followed by the active liquid crystal to
reach its steady state. For concreteness, henceforth we will restrict to the
case of normal anchoring.

Fig. \ref{dynamics400} show the dynamics followed by a system with
$L=40$ $\mu$m, which started from a (slightly perturbed)
spatially constant director field
with normal anchoring and no flow. 
The value of the activity parameter is one for which, for $L=10$ $\mu$m,
one ends up with states similar to those in Fig. 2. In the larger sample
this value of $\zeta$ is enough to lead to an out-of-plane director field
in steady state. This is in line with previous observations that
larger systems are more easily led into an active state \cite{joanny}.
The early time dynamics is
characterised by the generation of a lot of 
deformation in the director field, 
(reflected in the large spatial changes in
$n_{x,y,z}$), caused by the appearance of
spontaneous flow. Thermodynamic forces then set in to
reduce the number of `bands' or equivalently to reduce the
amount of splay and bend. Note also that transiently the director
field develops a significant asymmetry, so that it is twisted
unevenly, which gradually disappears upon approaching the
steady state.

Note that we have also simulated situations in which the activity
value is cyclically increased and decreased so that the initial
condition is different in the two halves of a 'cycle', to
compare the steady states found in the two cases (data not shown).
While with small to moderate values of the activity, for which the active
phase has just set in, we find no dependence on the initial conditions, 
this is not true as we go deeper in the active phase.
In this region we find there is some hysteresis, and that the steady state does
depend on initial condition for large value of $\zeta$ (i.e. deep
in the active phase). This is perhaps not surprising because
when subjected to a flow (in this case stimulated by the activity), 
liquid crystals and other viscoelastic gels may
exhibit rheochaotic (hence strongly initial
condition dependent) flow properties \cite{rheochaos}. It
would be interesting to systematically characterise the dependence of
steady state and of the dynamic pathway in active gels for large
$\zeta$, and determine whether, under appropriate conditions,
complex, or even chaotic, dynamics are observed. This is deferred to
future work.

\section{Conclusions and discussions}

In conclusions, we have presented a generalised lattice Boltzmann algorithm,
inspired by those previously used to solve the Beris-Edwards
equations of motion of a passive liquid crystal, to investigate numerically
the hydrodynamic properties of an active (flow-aligning) liquid crystalline
fluid in an effectively 1D geometry. We have studied two distinct
boundary conditions (one with the director field normally anchored at the
boundary plates and another one with conflicting anchoring leading to elastic
distortions even in the passive phase), and
systematically considered how the director and
flow fields in steady state evolve upon increasing the parameter
$\zeta$, which is a measure of the liquid crystal 'activity'.

The interplay of activity, which may introduce spontaneous flow in
steady state, and the elastic distortion caused by the fixed anchoring
at the boundaries, gives rise to an interesting phenomenology.
We show that our algorithm gives a non-equilibrium transition
from a passive to an active phase, and that this occurs with zero
threshold when there is conflicting anchoring at the boundaries.
Deep in the active phase, where semi-analytical methods are
typically unhelpful, we find further `transitions' between
qualitatively different steady states of flow and director field.
When normal anchoring is considered, we observe banded flow solutions
in which regions of shear-oriented filaments or director field
are separated by regions of bend-splay distortions (where the
order parameter decreases slightly), as well as states with
director field out of the $xz$ plane in steady state.

The equations of motion we have studied are expected to describe the coarse 
grained dynamics of actomyosin solutions or of microtubular networks with 
dyneins, when the actin fibers or the microtubules have a fixed orientation at
the surface (due e.g. to chemical or mechanical treatment of the surface as in
the csae of passive liquid crystal). We note that
our algorithm can be straightforwardly applied to 2D and 3D problems;
this will be pursued elsewhere.

\newpage

\begin{center}
{\bf Figure captions}
\end{center}

{{\bf Fig. 1:} Geometry used for the calculations described in the
text. The active gel is sandwiched between two
infinite plates, parallel to the $xz$ plane, lying at
$y=0$ and $y=L$. We consider (a) normal anchoring and (b) conflicting
anchoring  (leading to a hybrid aligned nematic cell for a passive liquid
crystal)}

{{\bf Fig. 2:} Director (left) and velocity (right) fields in
steady state for an active liquid crystal slab with normal anchoring at the
boundaries, and $\zeta$ equal to 0.0002 (solid line), 0.0004
(long-dashed line), 0.0006 (dot-dashed line), and 0.001 (dotted
line), respectively. Other parameters are as specified in the
text.}

{{\bf Fig. 3:} Director (left) and velocity (right) fields in
steady state for an active liquid crystal slab with normal
anchoring at the boundaries, and $\zeta$ equal to
0.002 (solid line), 0.003 (long-dashed line), 0.005 (dot-dashed line),
and 0.007 (dotted line) respectively. Other parameters are as specified in
the text.}

\begin{figure}
\centerline
{\includegraphics[width=10.cm]{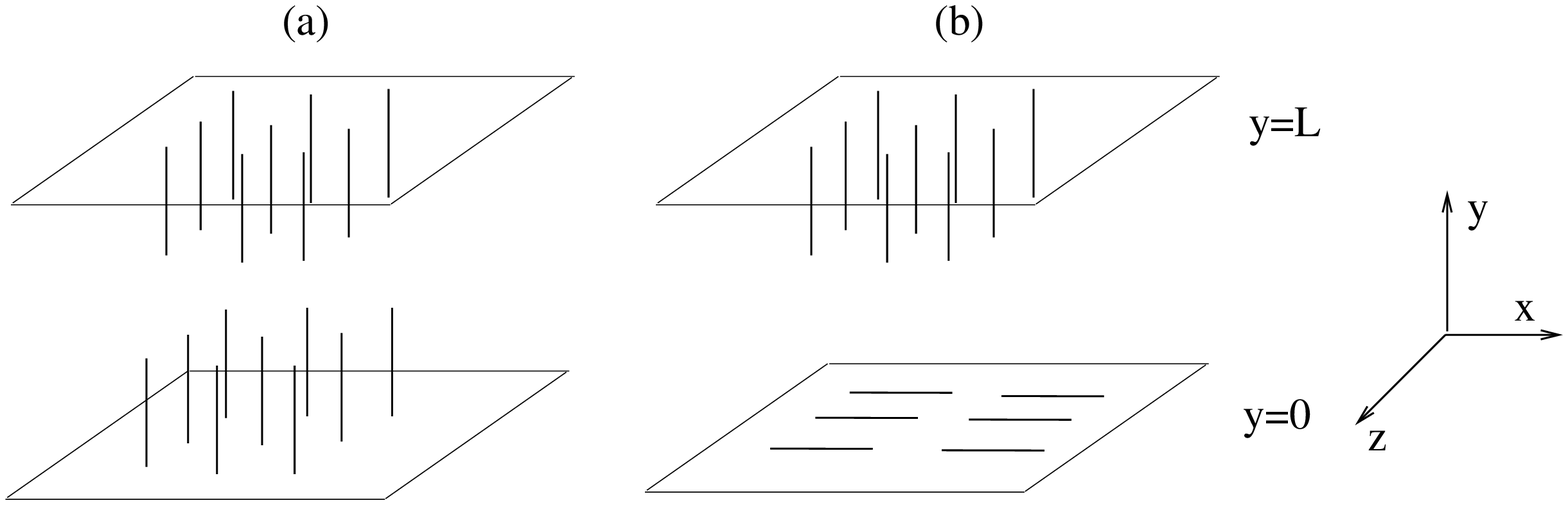}
}
\caption{Geometry used for the calculations described in the
text. The active gel is sandwiched between two
infinite plates, parallel to the $xz$ plane, lying at
$y=0$ and $y=L$. We consider (a) normal anchoring and (b) conflicting
anchoring  (leading to a hybrid aligned nematic cell for a passive liquid
crystal).\\[8.cm]}
\label{setup}
\end{figure}

\begin{figure}
\centerline {\includegraphics[width=8.cm]{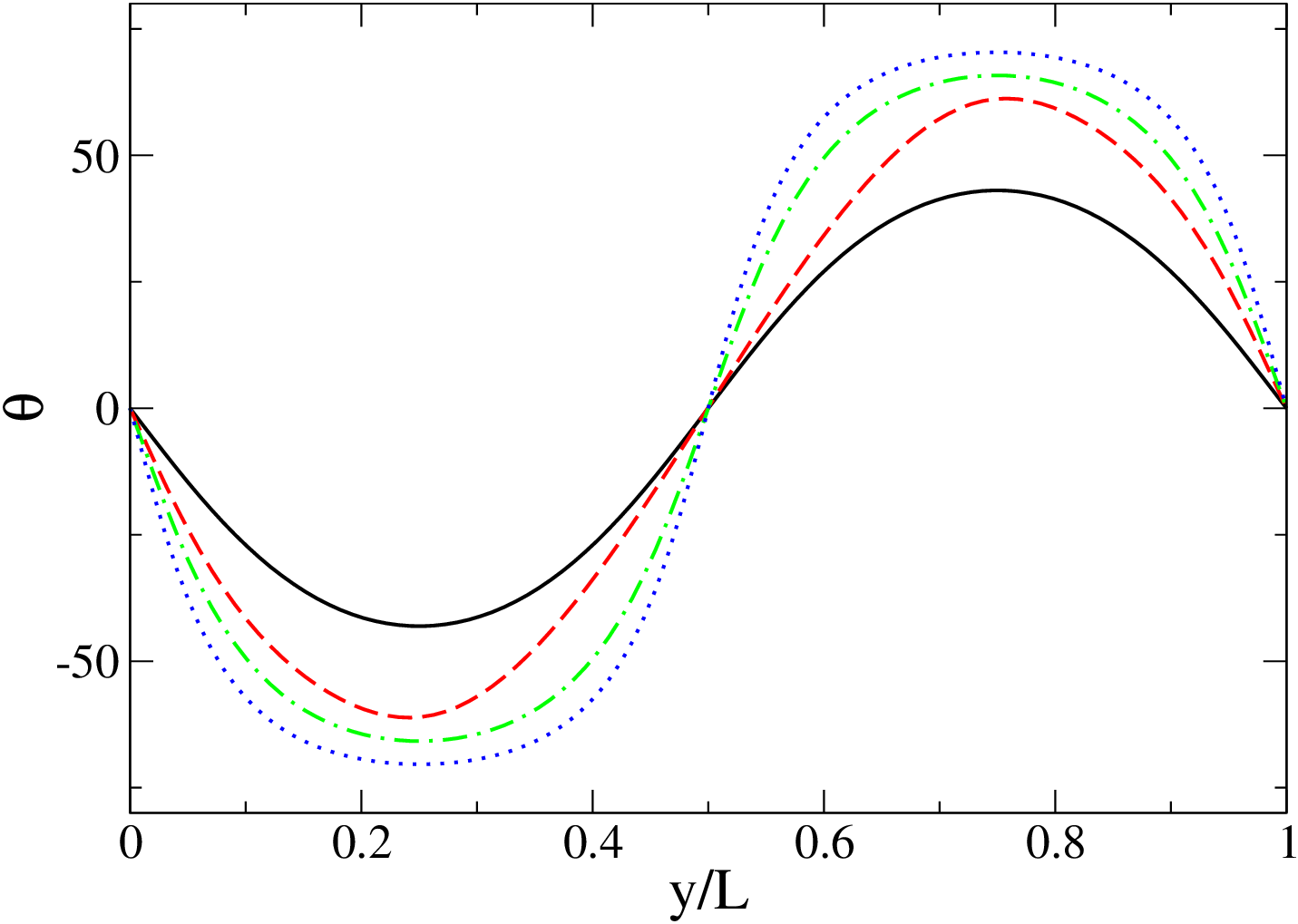} \qquad
\includegraphics[width=8.6cm]{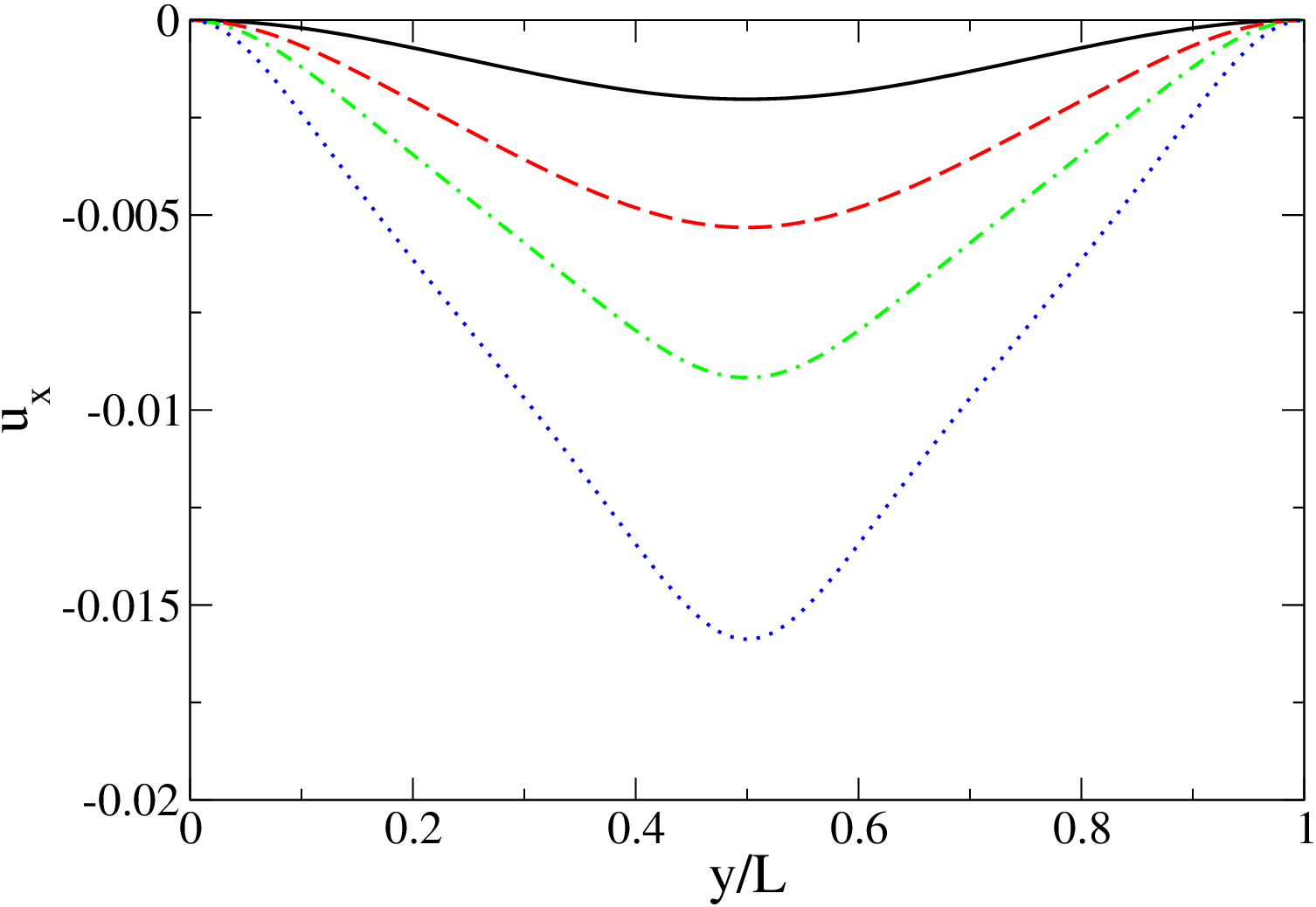}} 
\caption{Director (left) and velocity (right) fields in
steady state for an active liquid crystal  slab with normal anchoring at the
boundaries, and $\zeta$ equal to 0.0002 (solid line), 0.0004
(long-dashed line), 0.0006 (dot-dashed line), and 0.001 (dotted
line), respectively. Other parameters are as specified in the
text.\\[8.cm]} \label{N1}
\end{figure}

\begin{figure}
\centerline
{\includegraphics[width=8.cm]{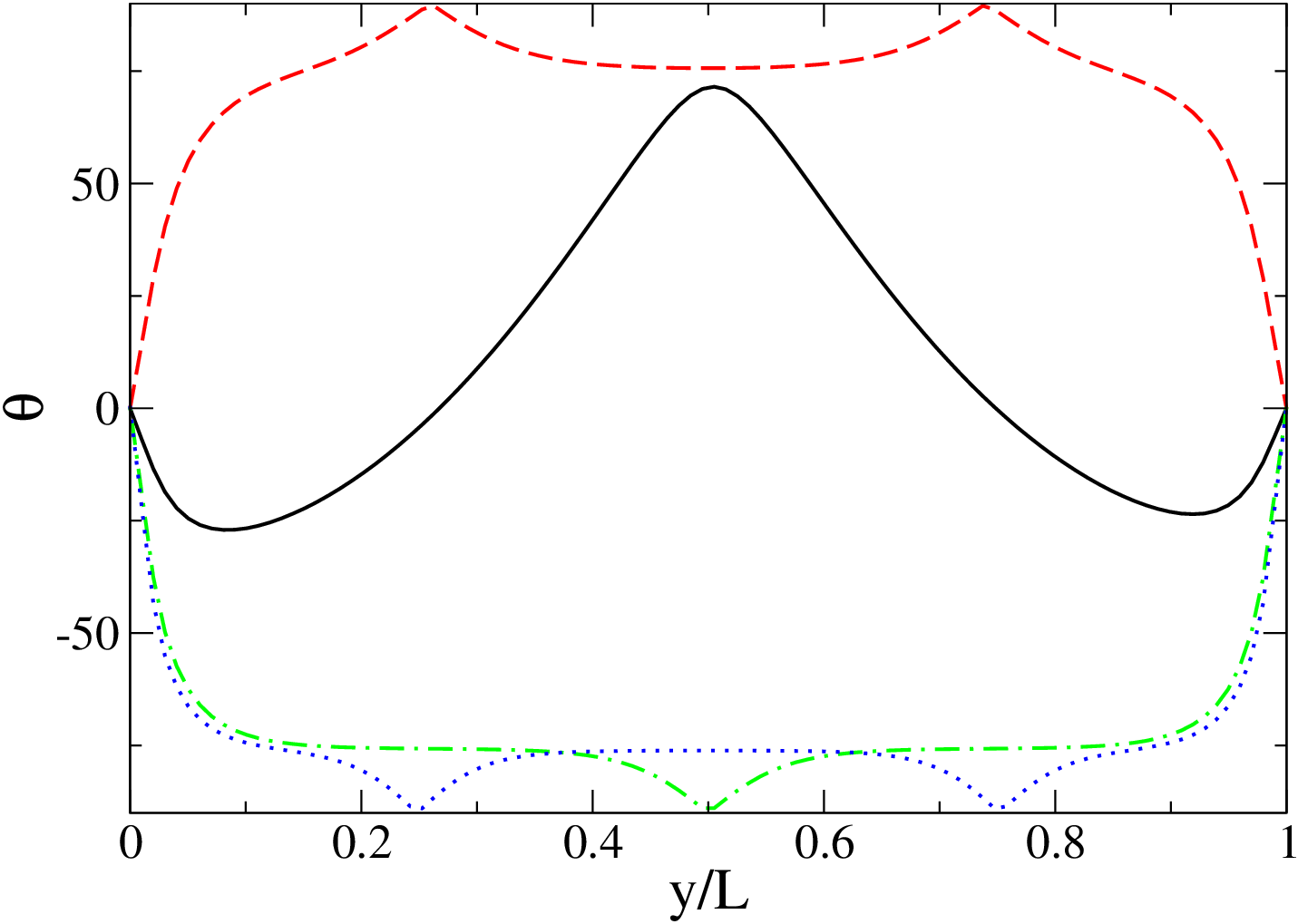} \qquad
\includegraphics[width=8.6cm]{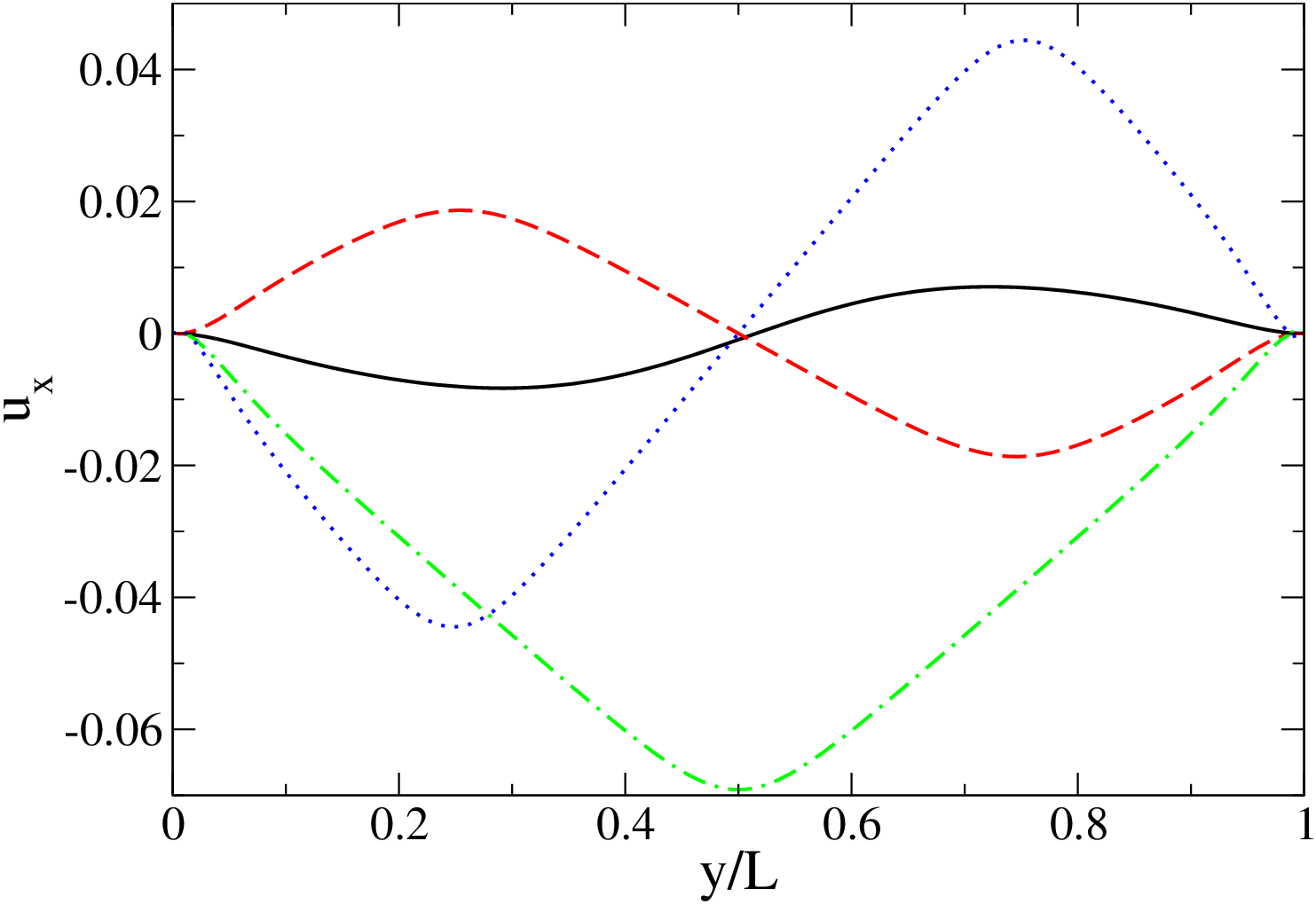}
}
\caption{Director (left) and velocity (right) fields in
steady state for an active liquid crystal slab with normal
anchoring at the boundaries, and $\zeta$ equal to
0.002 (solid line), 0.003 (long-dashed line), 0.005 (dot-dashed line),
and 0.007 (dotted line) respectively. Other parameters are as specified in
the text.\\[8.cm]}
\label{N2}
\end{figure}

\begin{figure}
\centerline
{\includegraphics[width=8.cm]{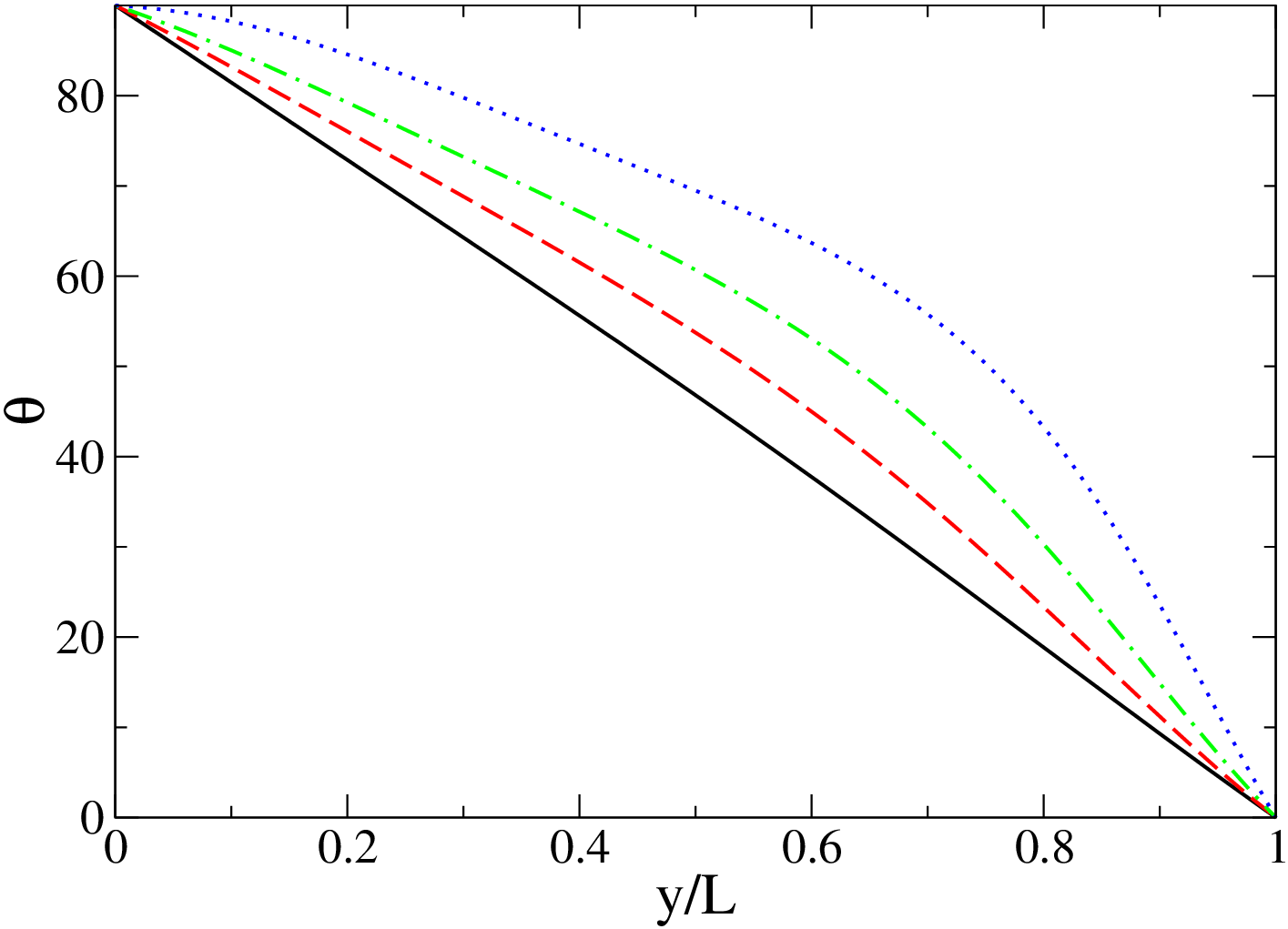} \qquad
\includegraphics[width=8.6cm]{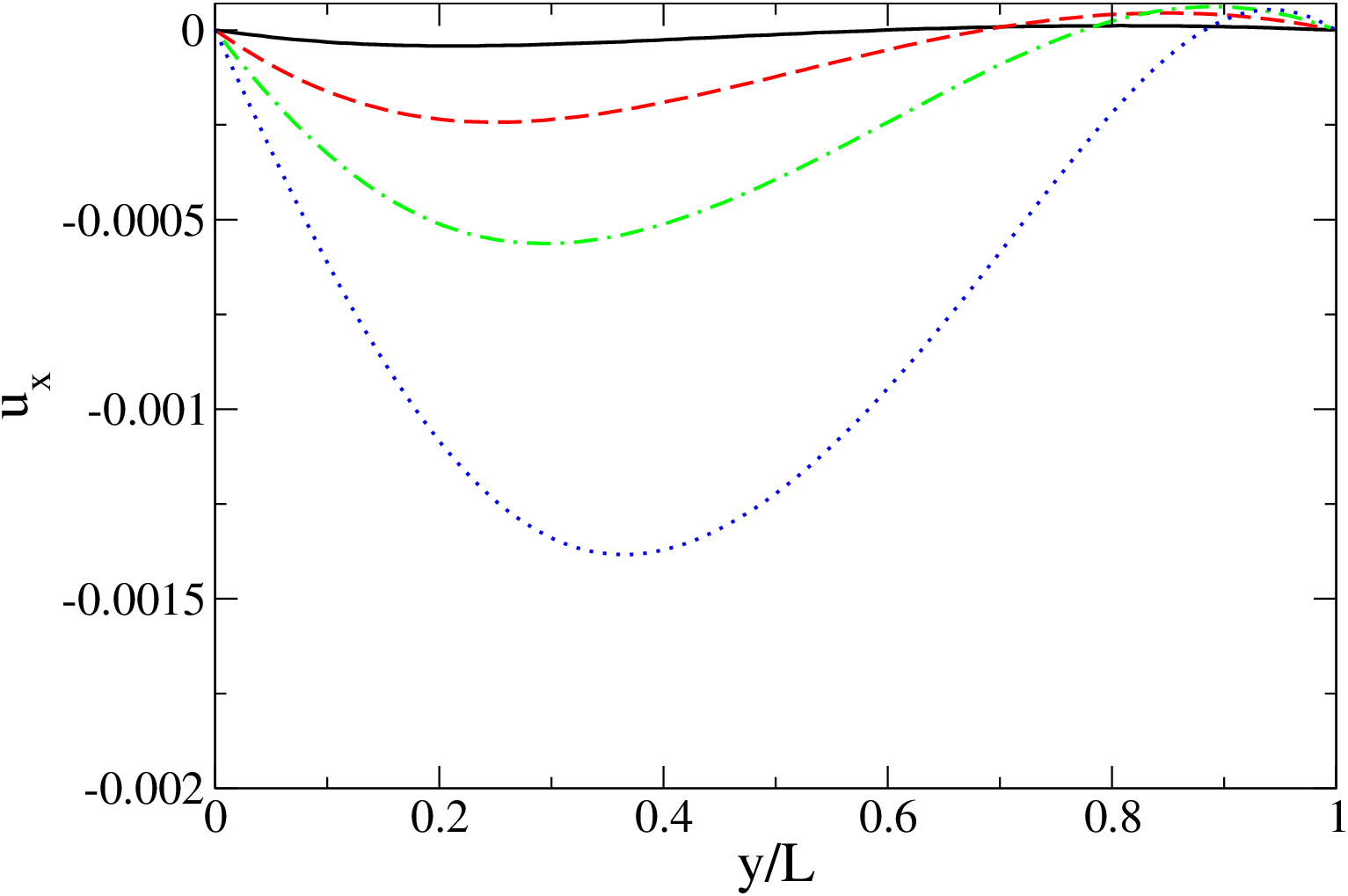}
}
\caption{Director (left) and velocity (right) fields in
steady state for an active liquid crystal slab with conflicting
anchoring at the boundaries (Fig. \ref{setup}, and $\zeta$ equal to
0.00001 (solid line), 0.00005 (long-dashed line), 0.0001 (dot-dashed line),
and 0.0002 (dotted line), respectively. Other parameters are as specified in
the text.\\[8.cm]}
\label{HAN1}
\end{figure}

{{\bf Fig. 4:} Director (left) and velocity (right) fields in
steady state for an active liquid crystal slab with conflicting
anchoring at the boundaries (Fig. \ref{setup}, and $\zeta$ equal to
0.00001 (solid line), 0.00005 (long-dashed line), 0.0001 (dot-dashed line),
and 0.0002 (dotted line), respectively. Other parameters are as specified in
the text.}

{{\bf Fig. 5:} Director (left) and velocity (right) fields in
steady state for an active liquid crystal slab with conflicting
anchoring at the boundaries (Fig. \ref{setup}, and $\zeta$ equal to
0.0004 (solid line), 0.0006 (long-dashed line), 0.0008 (dot-dashed line),
and 0.003 (dotted line), respectively. Other parameters are as specified in
the text.}

{{\bf Fig. 6:} Dynamic pathway to steady state
followed by a system with $\zeta=0.0008$, $L=400$,
and other parameters as in the previous section.
Solid, long dashed and dot-dashed lines correspond to
the $z$, $x$ and $y$ components of the director field respectively.}

\newpage

\newpage

\newpage

\begin{figure}
\centerline {\includegraphics[width=7.7cm]{HAN_pol2.eps}\qquad
\includegraphics[width=8.cm]{HAN_ux2.eps}
}
\caption{Director (left) and velocity (right) fields in
steady state for an active liquid crystal slab with conflicting
anchoring at the boundaries (Fig. \ref{setup}, and $\zeta$ equal to
0.0004 (solid line), 0.0006 (long-dashed line), 0.0008 (dot-dashed line),
and 0.003 (dotted line), respectively. Other parameters are as specified in
the text.\\[8.cm]}
\label{HAN2}
\end{figure}

\begin{figure}
\centerline
{\includegraphics[width=14.cm]{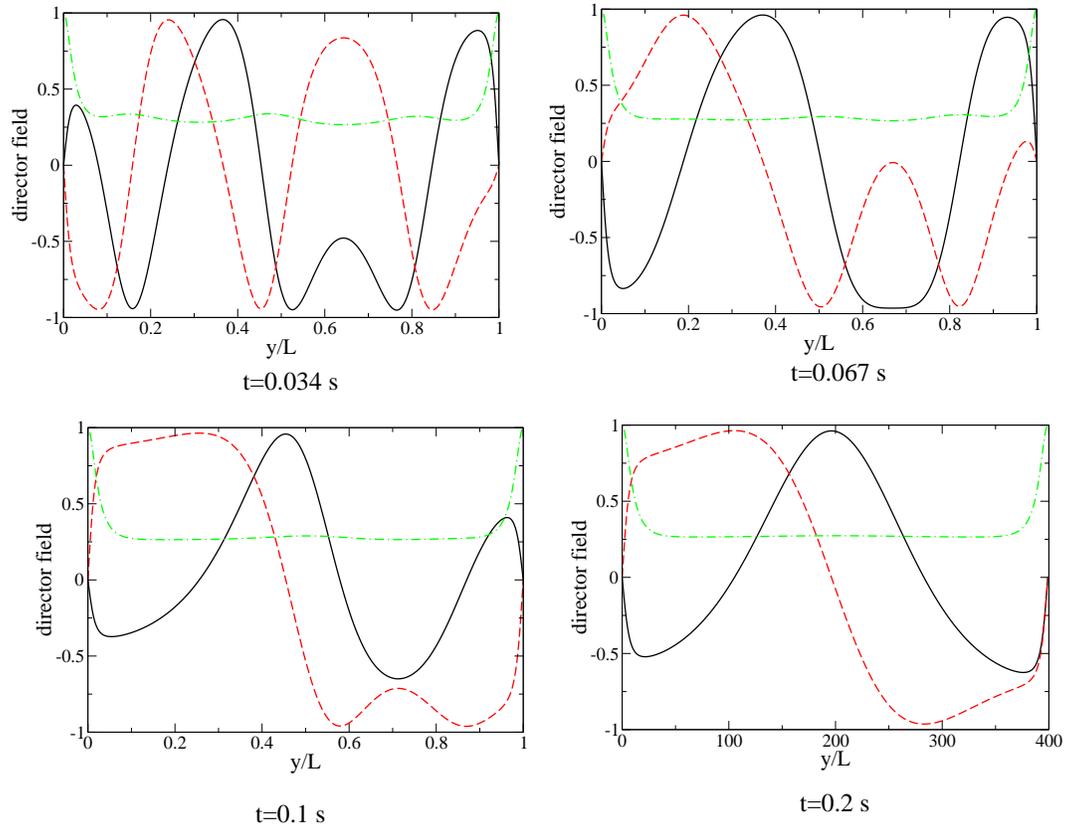}}
\caption{Dynamic pathway to steady state
followed by a system with $\zeta=0.0008$, $L=400$,
and other parameters as in the previous section.Solid, long dashed and 
dot-dashed lines correspond to
the $z$, $x$ and $y$ components of the director field respectively.}
\label{dynamics400}
\end{figure}

\end{document}